\documentclass[a4paper,11pt]{article}
\usepackage{jheppub} 

\usepackage{tkz-graph}
 \usepackage{comment}
\newcommand{\ivo}[1]{\textcolor{red}{#1}}
\begin{document}
\begin{flushright}
	LMU-ASC 01/25\\
\end{flushright}

\title{\boldmath 
Yang-Mills Theory From Super Moduli Space
}

\author[a]{Carlo Alberto Cremonini}
\author[a]{and Ivo Sachs}

% The "\note" macro will give a warning: "Ignoring empty anchor..."
% you can safely ignore it.

\affiliation[a]{Arnold-Sommerfeld-Center for Theoretical Physics, Ludwig-Maximilians-Universit\"at, M\"unchen,\\ Theresienstr. 37, D-80333 M\"unchen, Germany}

% e-mail addresses: one for each author, in the same order as the authors
\emailAdd{carlo.alberto.cremonini@gmail.com}

\emailAdd{ivo.sachs@physik.lmu.de}

\abstract{For the spinning superparticle we construct the pull-back of the world-line path integral to super moduli space in the Hamiltonian formulation. We describe the underlying geometric decomposition of super moduli space. Algebraically, this gives a realization of the cyclic complex. The resulting space-time action is classically equivalent to Yang-Mills theory up to boundary terms and additional non-local interactions.
}

\maketitle
\flushbottom

%\pagestyle{myplain}
%\tableofcontents
\section{Introduction}
The formulation of superstring field theory (SSFT) is obscured in two ways: One is related to the intricacies of super moduli space $\mathcal{M}$ (moduli space of super Riemann surfaces). This, in turn, complicates the geometric decomposition of $\mathcal{M}$, and therefore the understanding of the SSFT action as an integral form on $\mathcal{M}$ (see, e.g. \cite{Belopolsky:1997jz, Witten:2012bg}). The second concerns the construction of this integral form in terms of the world-sheet conformal field theory (CFT) of the superstring, which is complicated by the lack of a convenient CFT representation of the differentials on super moduli space. Usually, the approach followed is to split the integration over super moduli space by representing the odd integration by some Picture Changing Operators (PCOs) in the CFT and keeping the bosonic integration as a geometrical operation\footnote{There is an alternative and earlier construction in the ``large'' Hilbert space \cite{Berkovits:1995ab} which is not based on super moduli space. Furthermore, this theory can be mapped to the algebraic construction in \cite{Erler:2013xta} as shown in \cite{Erler:2015rra,Erler:2015uoa}.}. This can be done either in the the world-sheet approach  \cite{Sen:2015uaa} or in terms of algebraic vertices as maps on the string state space \cite{Erler:2013xta}. While this is consistent as long as $\mathcal{M}$ is split \cite{Donagi:2013dua,Ohmori:2017wtx, Wang:2022zad}, this hybrid description blurs the geometric formulation as an integral form\footnote{In the supergeometric context, PCOs are (co)chain maps between the complexes of differential and integral forms on a given supermanifold which are quasi-isomorphisms  \cite{Cremonini:2023ccv}. %(see e.g. \cite{Cremonini:2023ccv} for the introduction of PCOs in this context).
The map that lifts a differential form to an integral form is globally defined as it is related to the (canonical) embedding of the reduced manifold in the ambient supermanifold; on projected supermanifolds, the map in the other direction is related to the (non-canonical) projection of the supermanifold on its reduced counterpart. %On non-projected supermanifolds, the definition of this map is more subtle and still the subject of investigation.
} (see also \cite{Sen:2015hia}).

One may hope to gain insight into some of these issues by replacing the super Riemann surface with a super world-line, since for infinite string tension the world-sheet should reduce to a world-line. The odd part of the moduli space $\mathcal{M}$ is preserved by this reduction so that the intricacies added by the odd directions persist. On the other hand, the world-line sigma model is considerably simpler than the world-sheet CFT. In particular, its BRST quantization can be performed off-shell, thanks to the absence of anomalies. In this paper, we will then derive the space-time action implied by the spinning particle as an integral form, integrated over super moduli space, following the same prescription as for the superstring. However, our approach differs from the standard description in some ways. First, we use the Hamiltonian action for the world-line rather than the Lagrangian action. This makes, the relation between the canonical off-shell BRST quantization of the corresponding constraint system manifest%\footnote{While it is possible to start with the Hamiltonian world-sheet action for the string as well this is not easily related to the oscillator formulation that is usually considered there.}
. Furthermore, we consider two different formulations of the off-shell path-integral. 

The first is analogous to the standard embedding of off-shell states into the world-sheet by suitable conformal mappings. Following this path, we will see that the integral of the pull-back of the world-line correlators over $\mathcal{M}$ realizes the BV-extension of Yang-Mills theory in a non-standard formulation, with additional non-local interactions induced by massless higher form auxiliary fields. In string theory, these additional fields are suppressed by the string tension. Accordingly, this outcome is consistent with the low energy limit of string field theory \cite{Berkovits:2003ny}, although there is no space-time supersymmetry here. The world-line field theory we obtain is quartic. The quartic vertex is particularly interesting: Just like in string theory, the four-punctured world-line has a bosonic modulus $\tau$, and would thus lead to additional non-local interactions. However, as we will see the integral over the two odd coordinates in  $\mathcal{M}$ (corresponding to the two odd moduli due to the insertion of inserting $4$ vertex operators on the world-line) results in a total derivative w.r.t. $\tau$ such that the quartic vertex reduces to a pure boundary term in $\tau$, i.e. a contact term. This feature is also implicit in \cite{Ohmori:2017wtx}. Tracing this back to the geometry of $\mathcal{M}$ we are then able to obtain a geometric decomposition of $\mathcal{M}$.  

In the second approach, we treat the insertions in the world-line path integral as background fields. This is somewhat analogous to background independent string field theory (see, e.g. \cite{Witten:1992qy,Shatashvili:1993ps, Chiaffrino:2018jfy}) or the sigma-model approach to SFT (see, e.g. \cite{Tseytlin:2000mt, Kostelecky:1999mu, Ahmadain:2022tew}). In this formulation, it is less clear what the correct decomposition of the moduli space of a punctured world-line is, since the background fields need not be simply related to scattering states. If we add the additional requirement that the space-time functional of the fields should be local, we can argue that we can consider no more than four punctures, with the presence of the quartic term required by gauge invariance. What we then find is that the pull-back of this world-line correlator integrated over $\mathcal{M}$ has the interpretation of a quadratic action with a background field dependent BRST operator $Q(A)$. %Furthermore, nilpotence of $Q(A)$ implies again the Yang-Mills. 
This is the manifestly background-independent formulation of the world-line field theory. Background independence of SFT has previously been discussed in e.g. \cite{Sen:1993kb, Munster:2012gy,Sen:2017szq}.

The plan of the remaining sections is as follows: In section \ref{sec:2} we begin by describing the super geometry of the super world-line and, in particular, identify moduli and automorphisms. In section \ref{sec:perturbative_action} we construct the pull-back of 2-, 3-, and 4-point world-line correlators as top forms on the corresponding super moduli spaces and integrate them to obtain the space-time action in the standard approach, i.e. by inserting off-shell states at the punctures. We find that a quartic interaction is required both by geometry and gauge-invariance of the action. However, the full equations of motions are already implied at the cubic level, upon variation w.r.t. to the Lagrange multiplier fields. Here we also describe the geometric decomposition of $\mathcal{M}$. In section \ref{sec:background fields} we repeat this construction for the insertion of background fields, rather than states. We find the space-time functional has a rather different interpretation in terms of background-independent construction of the space-time BRST-differential acting on 1-particle states. The latter are well-defined on classical solutions of the background fields. 

\section{Super Moduli Space and BRST Quantization}\label{sec:2}
\subsection{Geometric description}\label{sec:geometricdescription}
Let us begin by recalling the description of super reparametrization invariance of the world-line and the resulting super moduli space $\mathcal{M}$. The discussion of the world-sheet superconformal transformations and world-line reparametrization proceeds in close analogy so that we may as well start with the former (see e.g. \cite{Witten:2012bg} for details and references) and emphasize the difference with the world-line when necessary. 
They are generated by the even and odd vector fields 
\begin{align}\label{vector fields}
    V_g&=g(z)\partial_z+\frac{g'(z)}{2}\theta\partial_\theta\,,\qquad
    v_f=f(z)(\partial_\theta-\theta\partial_z)\,,
\end{align}
under which the supercovariant derivative\footnote{Or equivalently a local generator of the rank $(0|1)$ sub-bundle defining super Riemann surfaces.}
\begin{align}
    D_\theta=\partial_\theta+\theta\partial_z
\end{align}
transforms as 
\begin{align}
    (\theta f'+\frac{g'(z)}{2})D_\theta\,.
\end{align}
In particular, $f=\epsilon$, independent of $z$, induces a global SUSY transformation while $g=const$ induces translations. Since we will be interested in classical string/world-line field theory, we may assume the vanishing genus of the world-sheet. Moreover, for the world-line (with $z\equiv s$) the automorphism group is generated by the (integral curves of) the vector fields that leave $D_\theta$ invariant\footnote{Since they preserve the gauge fixing, $e\equiv 1$ and $\chi=0$, of the Einbein and world-line gravitino.} which are just the global transformations above.  

To describe the moduli space we need to include punctures, where vertex operators $\mathcal{V}_i$ are inserted. Then the redundant reparametrizations are generated by vector fields (both even and odd) that vanish at those punctures. More precisely, the tangent space to the moduli space is generated by vector fields that don't vanish at a given puncture, modulo those that vanish instead (which we will refer to as gauge generators) and the global transformations described above. 

For concreteness, let us consider the world-line with 3 punctures with local coordinates $(s_i,\theta_i)$, $i=1,2,3$  around each puncture: 
\begin{equation}\label{fig:wl3}
\begin{gathered}\begin{tikzpicture}
\draw[very thick] (-3, 0) -- (3, 0);
\node at (-3, 0) {\textbullet};
\node at (0, 0) {\textbullet};
\node at (3, 0) {\textbullet};
\node at (-3, -0.4) {$\mathcal{V}_1(-\infty)$ };
\node at (3, -0.4) {$\mathcal{V}_3(\infty) $ };
\node at (0, -0.4) {$\mathcal{V}_2(s_0)$ };
\end{tikzpicture}\end{gathered}
\end{equation}
In the BRST quantization of the world-line/world-sheet, described in the next section, the Einbein $e$ (i.e., the world-sheet metric) is gauge fixed to a constant and the gravitino $\chi$ is gauge fixed to zero. The admissible super reparametrizations are those for which the transformation of $e$ and $\chi$ are non-singular. 

We first recall the situation of the upper half plane, relevant for the open superstring, and see how things are modified when restricting to the line. Let us consider the reduced upper half plane (i.e., the one where the odd coordinates are set to zero). There we have three globally defined even holomorphic vector fields, $\partial_z$,  $z\partial_z$ and $z^2\partial_z$. On the world-line, the constant even vector field $g=const$ translates  $s_0$ but leaves $s=\pm\infty$ invariant and represents the even automorphism of the line. The vector field with $g(s)=s$ does not move any of the punctures\footnote{As can be seen after mapping this point to zero with the super reparametrization $t=-\frac{1}{s}$, $\xi=\frac{\theta}{s}$ for which $\partial_s=t^2 \partial_t$ and $\partial_\theta= -t\partial_\xi$.} but $\delta e=e\dot g $ is well defined. It thus becomes part of the reparmetrizations that can be eliminated by gauge fixing. Finally, $g(s)=s^2$ is excluded since $\delta e=e\dot g $ diverges at $s=\pm\infty$.

Similarly, the odd vector field with $f=\epsilon=const$ acts trivially at $\pm \infty$ and represents an odd automorphism of the super world-line. $f(s)=\epsilon s$ translates the odd coordinates at $s=\pm \infty$ and furthermore $\{v_f,v_f\}=-2V_g$ with $g(s)=s^2$. Thus $f(s)=\epsilon s$ is excluded in accordance with $g(s)=s^2$.  In sum, there is no bosonic modulus and one fermionic modulus for the line with 3 punctures.

\subsection{Path-integral quantization}\label{sec:Path-integral quantization}

In the path integral, the reparametrization vector fields $V$ and $v$ (including the global transformations) are replaced by the parity reversed ghosts $c$ and $\gamma$, respectively. After gauge fixing of the Einbein $e$ and the gravitino $\chi$, the world-line path integral
\begin{align}\label{eq:pat}
    &\int\mathrm{d} g \;\mathrm{d} f\; \mathrm{D} [x ,p,\psi ,b ,c ,\gamma ,\beta ]\;e^{iI[\cdots]}\;\mathcal{V}_1(s_1)\mathcal{V}_2(s_2)\mathcal{V}_3(s_3)\cdots\,,
    \end{align}%\int[ \mathrm{d}\eta,\mathrm{d}(\mathrm{d}\eta)] 
 reduces to an integral form on the supermoduli space $\mathcal{M}$, if the action $I$ and the vertex oparators $\mathcal{V}_i$ are suitably chosen. Following \cite{Witten:2012bg}, this pull back is constructed through the BRST variation of the extended world-line action
 \begin{align}\label{eq:aext}
I[x,\psi,b,c,\beta,\gamma,e,\delta e,\chi,\delta\chi]=I[X,\psi,b,c,\beta,\gamma]+\int ( e P^2+\delta e \,b ) +\int ( \chi \mathbf{q}+\delta\chi \beta )\,,
 \end{align}
 where 
\begin{align}\label{eq:wlaction}
 I[x,p,\psi,b,c,\beta,\gamma]=    \int p\cdot\dot x-\frac{1}{2}p^2-\frac{i}{4}\psi\cdot\dot\psi -i\beta\dot\gamma+ib\dot c\,,
 \end{align}
 is invariant under the BRST-transformations generated by %\footnote{This implies $\langle\psi^\mu(t)\psi^\nu(t')\rangle=\text{sgn}(t-t')$ and $\langle p^\mu(t)x^\nu(t')\rangle=-\frac{i}{2}g^{\mu\nu}\text{sgn}(t-t')$ and thus $[p^\mu,x^\nu]= -ig^{\mu\nu}$ and $[\psi^\mu,\psi^\nu]=2g^{\mu\nu}$. Similarly,  $[\beta,\gamma]=1$ }
 \begin{align}
     Q=-cp\,^2+\gamma\psi\cdot p+b\gamma^2 \,.
 \end{align}
 From the measure \eqref{eq:wlaction} one derives \begin{align}\label{eq:two-points}
     \langle\psi^\mu(t)\psi^\nu(t')\rangle=\text{sgn}(t-t')\,,\quad\text{and}\quad \langle p^\mu(t)x^\nu(t')\rangle=-\frac{i}{2}g^{\mu\nu}\text{sgn}(t-t')
 \end{align} 
 and thus 
 \begin{align}
     [p^\mu,x^\nu]= -ig^{\mu\nu}\,\quad \text{and}\quad [\psi^\mu,\psi^\nu]=2g^{\mu\nu}\,.
 \end{align}
 Similarly, $[\beta,\gamma]=1$. For $\mathbf{q}=\psi\cdot p+2b\gamma$ the invariance of \eqref{eq:aext} follows if we define
 \begin{align}
     \delta_{BRST}e=\delta e\,,\quad  \delta_{BRST}\delta e= 0
 \end{align}
 and similarly for $\chi$ and $\delta\chi$. Then, setting 
 \begin{align}
  e=i\frac{\partial g(s,\tau)}{\partial s} \,,\qquad \delta e =i\frac{\partial (\partial_\tau g)(s,\tau)}{\partial s}\mathrm{d}\tau\,,\qquad \chi=i\frac{\partial f(s, \eta)}{\partial s}  \,,\qquad  \delta\chi=i\frac{\partial (\partial_\eta f(s, \eta))}{\partial s}\mathrm{d}\eta  \,,%= \frac{\partial}{\partial s}\frac{\partial  \chi}{\partial\eta}\mathrm{d}\eta
 \end{align}
 where $\tau$ and $\eta$ are local even and odd coordinates on $\mathcal{M}$ parametrizing the moduli. % \eqref{eq:pat} defines a form on $\mathcal{M}$.
Integration over $\mathcal{M}$ then yields a functional of the space-time fields parametrizing the vertex operators $\mathcal{V}_i$.  

Before evaluating this pull back, let us describe the vertex operators $\mathcal{V}$. We will use the notation $\mathcal{V} = V ( \psi , c , \gamma , \beta ) \rho$, where $\rho$ indicates the choice of the representation of the Weyl algebra defined by the $\beta , \gamma$ system, which could be given by either polynomials in $\beta$ (in that case $\rho = \delta ( \gamma )$) or polynomials in $\gamma$ (in that case $\rho = \mathbf{1}_\gamma$). $\delta ( \gamma)$ selects the representation whose ground state is annihilated by the multiplicative action of $\gamma$ and whose states are generated by acting with $\beta$, which is realised as a derivative operator. This corresponds to the choice of the ``integral form'' representation of the $\beta, \gamma$ system, see \cite{Witten:2012bg} for an introduction to the topic with contact to field and string theory. $\mathbf{1}_\gamma$ selects the representation whose ground state is annihilated by the action of $\beta$ and whose states are generated by acting with $\gamma$. This corresponds to the choice of ``superforms'' (or simply differential forms) representation of the $\beta,\gamma$ system. Let us start by considering a polynomial of odd parity and ghost degree 0,
\begin{align}
 V=& A + c A^* + \beta C + c \beta B + \beta^2 D + c \beta^2 E + \ldots \nonumber \\
 = & A^{[1]}+A^{[3]} + c ( A^{*[1]} + A^{*[3]} ) +\beta (C^{[0]}+ C^{[2]}+C^{[4]}) +c\beta( B^{[0]}+B^{[2]}+B^{[4]})  \nonumber\\
 \label{picture-1vertex} + & \beta^2 (D^{[1]}+D^{[3]}) + c\beta^2(E^{[1]}+E^{[3]}) +\cdots
\end{align}
(left) acting on $\mathbf{1}_x\cdot\delta(\gamma)$, where $\mathbf{1}_x$ is the ground state annihilated by the left action of the momentum operator $p$, so that the vertex is given by $\mathcal{V} = V \mathbf{1}_x \cdot \delta ( \gamma )$. Here, $ A^{[k]}=A_{\mu_1\cdots\mu_k}\psi^{\mu_1} \cdots \psi^{\mu_k}$ and $B^{[k]}$ are of ghost degree zero, $C^{[k]}$ and  $E^{[k]}$ of ghost degree 1, $D^{[k]}$, of ghost degree 2, is a ghost for ghost etc. In this paper we assume space-time to be flat and four dimensional. Both restrictions can be lifted without complications since there is no conformal anomaly to worry about. Also one could consider a ghost number zero multiplet of \textit{even} parity instead. Our choice is motivated because it is the same class as Yang-Mills theory.

Acting with $Q$ on $\mathcal{V}$ identifies the linearized equations of motion and the gauge symmetries\footnote{The BRST operator is acting on the left as the vertex is intended as a \emph{state}, i.e., we have specified a ground state.} 
\begin{align}
    Q \mathcal{V} & = \Big[ c \left[ \Box A^{[1]} + \Box A^{[3]} - i \mathrm{d} B^{[0]} - i \mathrm{d} B^{[2]} - i \delta B^{[2]}  - i\delta B^{[4]} \right] + \nonumber\\
    & + \left[ + i \mathrm{d} C^{[0]} + i \mathrm{d} C^{[2]} + i \delta C^{[2]} + i  \delta C^{[4]} + 2 E^{[1]} + 2 E^{[3]} \right] + \\
    & + c \beta \left[ \Box C^{[0]} + \Box C^{[2]} + \Box C^{[4]} - 2 i \mathrm{d} E^{[1]} - 2 i \delta E^{[1]} - 2 i \mathrm{d} E^{[3]} - 2 i \delta E^{[3]} \right] +\cdots \Big] \mathbf{1}_x \cdot \delta ( \gamma ) \ . \nonumber
\end{align}
Thus, the equations of motion read 
\begin{align}
    \Box A^{[1]} - i \mathrm{d} B^{[0]}- i \delta B^{[2]} = 0 \ , \nonumber\\
    \Box A^{[3]} - i \mathrm{d} B^{[2]} - i \delta B^{[4]} = 0 \ .
\end{align}
The gauge transformations read
\begin{align}\label{gauge N=1}
    \delta_g A^{[1]} & = i \mathrm{d} C^{[0]} + i \delta C^{[2]} + 2 E^{[1]} \ , \nonumber\\
    \delta_g A^{[3]} & = i \mathrm{d} C^{[2]} + i \delta C^{[4]} + 2 E^{[3]} \ , \nonumber\\
    \delta_g B^{[0]} & = \Box C^{[0]} - 2 i \delta E^{[1]} = - i \delta \left( \delta_g A^{[1]} \right) \ , \nonumber\\
    \delta_g B^{[2]} & = \Box C^{[2]} - 2 i \mathrm{d} E^{[1]} - 2 i \delta E^{[3]} = - i \mathrm{d} \left( \delta_g A^{[1]} \right) - i \delta \left( \delta_g A^{[3]} \right) \ , \nonumber\\
    \delta_g B^{[4]} & = \Box C^{[4]} - 2 i \mathrm{d} E^{[3]} = - i \mathrm{d} \left( \delta_g A^{[3]} \right) \ .
\end{align}
Let us count the degrees of freedom of the system in virtue of the gauge symmetries \eqref{gauge N=1}. For simplicity, let us go on-shell, i.e., $\Box = 0$. We can use the terms $E^{[1]}$ and $E^{[3]}$ to fix completely $B^{[0]}$, $B^{[2]}$ and $B^{[4]}$. The gauge parameters $C^{[0]}$ and $C^{[4]}$ remove one degree of freedom from $A^{[1]}$ and $A^{[3]}$, respectively, and the same with $C^{[2]}$. This leads to 2+2 degrees of freedom encoded in the remaining components of $A^{([1]}$ and $A^{[3]}$. Below, we will see that in the kinetic term obtained from the path integral, the vertices appear in the $\delta (\gamma)$ representation as well as in the ($\gamma$-) polynomial representation; we choose the vertices in the $\gamma$-polynomial representation to be of the form
\begin{align}
    \mathcal{V} = \left[ A + c A^* \right] \mathbf{1}_x \cdot \mathbf{1}_\gamma = \left[A^{[1]} + A^{[3]} + c ( A^{*[1]} + A^{*[3]} ) \right] \mathbf{1}_x \cdot \mathbf{1}_\gamma \ .
\end{align}
This form is non-canonical and reflects the fact that we are setting all the antifields (but $A^*$) to zero, thus spoiling the non-degeneracy of the symplectic form defined by the path integration. We comment more exhaustively on this in the appendix \ref{sec:Symplectic detour} also describing the completion of the theory with the other antifields. The $Q$-closure of the vertex in the polynomial representation implies the extra relations
\begin{align}\label{eq:p0close}
    & A^{*[1]} = 0 = A^{*[3]} \ , \nonumber\\
    & d A^{[1]} + \delta A^{[3]} = 0 \ ,
\end{align}
and thus we are left with 2 d.o.f. in total. So far, the formulation is symmetric in $A^{[1]}$ and $A^{[3]}$. We could take either of them to be free and the other constrained by \eqref{eq:p0close}. Below we will take $A^{[3]}$ to be the auxiliary field. 

\section{Perturbative space-time action}\label{sec:perturbative_action}
In this section, we follow the standard approach for constructing a string/world-line field theory \cite{Zwiebach:1993ie} by inserting states at the punctures of the world-line as in fig. \ref{fig:sa1}
\begin{equation}\label{fig:sa1}
\begin{gathered}\begin{tikzpicture}
\draw[very thick] (-3, 0) -- (3, 0);
\node at (-3, 0) {\textbullet};
\node at (0, 0) {\textbullet};
\node at (1, 0) {\textbullet};
\node at (3, 0) {\textbullet};
\node at (-3, -0.4) {$\mathcal{V}_1(-\infty)$ };
\node at (3, -0.4) {$\mathcal{V}_4(\infty) $ };
\node at (1, -0.4) {$\mathcal{V}_3(\epsilon) $ };
\node at (0, -0.4) {$\mathcal{V}_2(0)$ };
\node at (0, 1) {\textbullet};
\node at (0, 1.5) {$\mathcal{V}_i$ };
\draw[thick, dashed, -latex](0, 1) -- (3, 0.1);
\draw[thick,dashed,-latex] (0, 1) -- (0, 0.1);
\draw[thick,dashed,,-latex] (0, 1) -- (1, 0.1);
\draw[thick,dashed, ,-latex] (0, 1) -- (-3, 0.1);
\end{tikzpicture}
\end{gathered} \ .
\end{equation}
The arrows indicate the mapping of the state $\mathcal{V}_j$ to the puncture at $s=s_j$ through
\begin{align}
    \mathcal{V}_j(s_{j})&=e^{is_{j}H} V(x,\psi,\beta,c) \mathbf{1}_x \cdot \delta(\gamma) \;e^{-is_{j}H}
\end{align}
With this prescription, the actual value of $s_j$ is not important since the propagation along the world-line implicit in the path integral cancels any ambiguity in $s_j$. Since in this section we will always consider the insertion of states (that is, we are always assuming the ground state $\mathbf{1}_x$ at every puncture), in order to avoid pedantic notations we will not specify this ground state for the rest of the section. In section \ref{sec:background fields}, we will assume \emph{no ground state} for $x,p$.

\subsection{Quadratic Action}
For the kinetic term, which is quadratic in $\mathcal{V}$, there are no moduli, thus $\mathcal{M}$ is a point.  We then choose 
\begin{align}\label{eq:V123} 
 \mathcal{V}(s_{-\infty})&=e^{is_{-\infty}H} V(x,\psi,\beta,c)\mathrm{1}_\gamma \;e^{-is_{-\infty}H}\nonumber\\
%        \mathcal{V}_2(s_0,0)&=e^{i(s_0)H}c(s_0) V_2(x(s_0),\psi(s_0),\beta(s_0))\delta(\gamma(s_0))e^{-i(s_{-\infty})H}\nonumber\\
        \mathcal{V}(s_{\infty}),&=e^{is_{\infty}H} V(x,\psi,\beta)\delta(\gamma)e^{-is_{-\infty}H}\,,
\end{align}
where $\mathbf{1}_\gamma = \mathbf{1}_{\gamma(s_{- \infty})}$ and $\delta ( \gamma ) = \delta ( \gamma ( s_\infty ) )$. The first vertex is the picture zero ghost vacuum, while the second one realizes the picture $-1$ ghost vacuum that accounts for the $\gamma$-ghost zero mode in the path integral.  

Inserting $\mathcal{V}_1 = \mathcal{V} ( s_{- \infty})$, $\mathcal{V}_2 = Q \mathcal{V} ( s_{\infty})$ and $\mathcal{V}_3 = \mathcal{V}_4 = \ldots = 0$ in \eqref{eq:pat} we get\footnote{It holds that $\langle Q a ,b\rangle =(-1)^a\langle a ,Q b\rangle $.}
\begin{align}\label{eq:S2n}
    S^{(2)}&=\langle (A+cA^*) QV\delta(\gamma)\rangle\nonumber\\
    &=\int d^4x \left[\langle A,-\Box A+i\mathrm{d}B +i\delta B)+\langle A^*,2E+ i\mathrm{d}C + i\delta C\rangle\right ]\,,
\end{align}
where $\langle\cdots\rangle$ denotes the expectation value w.r.t. the measure defined in \eqref{eq:wlaction}% and $(\cdot,\cdot)$ is the inner product on $p$-forms\footnote{We take $\Box=\mathrm{d}\delta+ \delta \mathrm{d} $ and $(\alpha,\mathrm{d}\beta)=-(\delta\alpha,\beta)$} \textcolor{blue}{We should pay attention to this footnote!}
. Variation w.r.t. $B^{[0]}$ implies $\delta A^{(1)} = 0$. Furthermore, the expected gauge transformation of $A$ is recovered from the variation w.r.t. $A^*$. Finally, variation w.r.t. $B^{[2]}$ enforces $\mathrm{d} A^{[1]} +\delta A^{[3]}=0$. This is related to the asymmetry between $\mathcal{V}(s_{-\infty})$ and $\mathcal{V}(s_{\infty})$ in the quadratic action (since they are defined in different representations of the $\beta,\gamma$ system). 

\subsection{Cubic term}
Let us now turn to the cubic interaction. According to the discussion in section \ref{sec:geometricdescription} there is one odd modulus which we parametrize by the odd function $f(s)=\eta \chi(s)$ where $\chi(s)$ is some characteristic function whose support contains $s=0$. Thus, $\mathcal{M}$ is a ``fat point'' of dimension $(0|1)$. Then
\begin{align}%\mathrm{d}e_{gf}
&S^{(3)}=\int_\mathcal{M} \mathrm{d} f \int \mathrm{D}[x,  p,\psi,b,c,\gamma,\beta ]\;e^{iI[\cdots]}\;\mathcal{V}_1(s_{-\infty})\mathcal{V}_2(s_0)\mathcal{V}_3(s_{\infty})
\end{align}
where now %[[PERHAPS WE SHOULD ABSORB $c(s_{-\infty})$ in $\mathcal{V}_1(s_{-\infty},0)$ IN THE SAME WAY AS $\delta(\gamma)$]]
\begin{align}\label{eq:V123t} 
 \mathcal{V}_1(s_{-\infty})&=e^{is_{-\infty}H} V_1(x,\psi)\;c(s_{-\infty})\;\mathbf{1}_\gamma\;e^{-is_{-\infty}H}\nonumber\\
       \mathcal{V}_2(s_0)&=e^{is_0H} V_2(x,\psi,\beta,c)\delta(\gamma(s_0))e^{-is_{0}H}\nonumber\\
        \mathcal{V}_3(s_{\infty})&=e^{is_{\infty}H} V_3(x,\psi,\beta,c)\delta(\gamma(s_{\infty}))e^{-is_{\infty}H}\,.
\end{align}
Indeed, according to section \ref{sec:geometricdescription}, we have to saturate one even zero-mode with the insertion of $c$, one odd zero mode and one odd modulus with the insertion of two vertices in the integral form representation. We choose to insert $c$ at $s_{- \infty}$ and $\delta ( \gamma )$ at $s_0$ and $s_\infty$. The $\beta,\gamma$-system provides the Jacobian for the change of variables from $\chi$ to $f$ and $\mathrm{d} f$ represents the integration over the ``would be'' gauge mode in the absence of the puncture, which now represents an odd modulus. Then,
\begin{align}\label{eq:S3mod}%\mathrm{d}e_{gf}
    &S^{(3)}=-\int_\mathcal{M} [\mathrm{d}\eta \;\mathrm{d}(\mathrm{d}\eta)]\;f' \;\mathrm{d}c\,\mathrm{d}\gamma \;\langle \mathcal{V}_1({s_{-\infty}})  e^{iF_{s_0}} \mathcal{V}_2(s_{0}) e^{-iF_{s_0}}  \mathcal{V}_3({s_{\infty}}) \rangle
\end{align}
where $\langle \cdot,\cdot\rangle$ stands for the expectation value of the $(x,p,\psi)$ system and $\mathrm{d}c\,\mathrm{d}\gamma$ denotes the integration over the $c$- and $\gamma$ zero-modes which are saturated by the insertion of $c(s_{-\infty})$ and $\delta(\gamma(s_{\infty}))$ respectively. Furthermore, $F_{s}= if (p \cdot \psi  + 2b \gamma )+i\partial_\eta f  \mathrm{d}\eta  \beta $ evaluated at $s$, stands for the boundary value of the two rightmost integrands in \eqref{eq:aext} upon time evolution between the punctures. Here we used the fact that $\mathbf{q}, H$ and $\beta$ are conserved under time evolution. On the one hand, the $\mathrm{d}(\mathrm{d}\eta)$ part of the measure $[\mathrm{d}\eta \;\mathrm{d}(\mathrm{d}\eta)]$ is necessary to cancel the Jacobian $f':=\partial_{\eta}f(0)$ from $f$ to $\eta$. On the other hand, it is also the natural invariant measure (or Berezinian) for integration on supermanifolds\footnote{In our context it can be viewed as the analogue of a Faddeev-Popov ghost for pulling $\mathrm{d}f$ back to $\mathcal{M}$.} (see again \cite{Witten:2012bh}). Using the Baker-Campbell-Hausdorff formula we get \begin{align}
    e^{iF}=e^{-fp\cdot \psi}e^{- f 2 b\gamma}  e^{- \mathfrak{f} \beta} e^{f \mathfrak{f}b}  \,,
 \end{align} 
where $\mathfrak{f}= f'\mathrm{d}\eta$.  Thus 
\begin{align}\label{eq:fvdgf}
     &e^{iF} V\delta(\gamma) e^{-iF}=  V\delta(\gamma-\mathfrak{f})+if(\mathrm{d A}+i\{ p,\cdot A\})\delta(\gamma-\mathfrak{f})-2f(-\frac{\mathfrak{f}}{2}+\gamma)A^* \delta(\gamma-\mathfrak{f})\\&-2f( -\frac{\mathfrak{f}}{2}+\gamma)B\delta'(\gamma-\mathfrak{f})
     -ifc(\mathrm{d B}+\delta B +iB\cdot p)\delta'(\gamma-\mathfrak{f})+E\delta^{(2)}(\gamma-\mathfrak{f}) +F\delta^{(3)}(\gamma-\mathfrak{f}) +\cdots \nonumber
\end{align}
which we substitute in \eqref{eq:S3mod}. The terms proportional to $c$ may be straightforwardly neglected, as they vanish because of the insertion of a $c$ in $\mathcal{V} ( s _{- \infty} )$. Integration over the $\gamma$- zero mode amounts to setting $\gamma=0$\footnote{Using $x\delta'(x)\sim-
\delta(x)$}. After integration over $\mathrm{d}(\mathrm{d}\eta)$, $\mathrm{d}c$ and finally $\mathrm{d}\eta$ we are left with
\begin{align}\label{eq:S3n}
    S^{(3)}=i\int d^4x \left[\langle  A,\mathrm{d}A +i\{ p,\cdot A\} -i B,A\rangle - i \langle A,A^*, C\rangle\right]\,.
\end{align}
Again $\langle a,b,c\rangle$ stands for the contraction of all $\psi^\mu$'s with sign$(s)$-functions\footnote{The conventions are $\delta A^{[1]} = \partial_\mu A^\mu$, $\Box =d\delta+\delta d$.}.

At this point it is important to remember that we inserted a \textit{state} at $s=0$, that is, the middle entry in \eqref{eq:S3n} acts on $\mathbf{1}_x$. This means that $\lbrace p , A \rbrace = p A = - i \delta A$. We represent this pictorially by the graph 
\begin{equation}\label{fig:wl3s}
\begin{gathered}\begin{tikzpicture}
\draw[very thick] (-3, 0) -- (3, 0);
\node at (-3, 0) {\textbullet};
\node at (0, 0) {\textbullet};
\node at (3, 0) {\textbullet};
\node at (-3, -0.4) {$\mathcal{V}_1(-\infty)$ };
\node at (0, -0.4) {$s_0$ };
\node at (3, -0.4) {$\mathcal{V}_3(\infty) $ };
\node at (1, 1.6) {$\mathcal{V}_2$ };
\draw[very thick] (0, 0) -- (1, 1);
\node at (1, 1) {\textbullet};
\end{tikzpicture}\end{gathered} \ .
\end{equation}
to emphasize that $p$ does not act on $\mathcal{V}_3(\infty) $. By consequence, \eqref{eq:S3n} is equivalent to
\begin{align}\label{eq:S3ns}
    S^{(3)}=i\int d^4x \left(\langle  A,\mathrm{d}A + \delta  A -i B,A\rangle - i \langle A,A^*, C\rangle\right)\,.
\end{align}
In order to describe this dynamical system we focus on the physical fields (recall that by imposing the e.o.m. of $E$ one sets $A^* = 0$): 
\begin{align}\label{eq:19-12-1}
  S=& S^{(2)}+S^{(3)} =\langle A,-\Box A+ i\mathrm{d} B +i\delta B\rangle+i\langle A, \mathrm{d} A +\delta A-i B,A\rangle\nonumber\\
    =& -\langle \mathrm{d} A, \mathrm{d} A \rangle -\langle \delta A, \delta A\rangle +i\langle \mathrm{d} A, B\rangle +i \langle \delta A , B\rangle+i\langle A, \mathrm{d} A +\delta A-i B,A\rangle\,.
\end{align}
By varying w.r.t. $B^{[0]}, B^{[4]}$ and $B^{[2]}$ respectively, we obtain
\begin{align}\label{eq:19-12-2}
i\delta A^{[1]}+\langle A, A\rangle&=0\nonumber\\
\langle\cdot, i \mathrm{d} A^{[3]}\rangle +\langle\cdot, A^{[3]}, A^{[3]}\rangle&=0\nonumber\\
\langle\cdot, i \mathrm{d} A^{[1]}\rangle +\langle\cdot,i\delta A^{[3]}\rangle +\langle\cdot, A,A\rangle &=0\,.
%A^{[3]}\rangle +\langle\cdot,\stackrel{=\frac{1}{2}[A^{[1]}_\mu,A^{[1]}_\nu]\psi^\mu\psi^\nu \rangle +\cdots} {A,A\rangle} =0\,.
\end{align}
Note that the 3rd line in \eqref{eq:19-12-2} already implies the full, non-linear Yang-Mills equation for 
\begin{align}
F_{\mu\nu} =[D_\mu,D_\nu]\,, \quad\text{with}\quad D_\alpha=\partial_\alpha -i[A^{[1]}_\alpha,\;\;]\,.
    \end{align}
This can be seen by acting with the adjoint de Rham differential $\delta$ on the third line, which implies 
\begin{align}
    D^\alpha F_{\alpha\mu}+\cdots=0\,, 
    \end{align}
where $\cdots$ stands for non-local corrections coming from terms without derivatives of the auxiliary $3$-form $A^{[3]}$. 

\begin{comment}
\textcolor{blue}{Remove paragraph} Upon substitution of \eqref{eq:19-12-2} into \eqref{eq:19-12-1} this gives we find in turn (using $\delta A^{[3]}\rangle =-\mathrm{d} A^{[1]}\rangle +i A,A\rangle$)
\begin{align}\label{eq:Sminusinfty}
 S^{(2)}+S^{(3)}=&2\langle \mathrm{d} A^{[1]}, \delta A^{[3]}\rangle\\
=& -2\langle \mathrm{d} A^{[1]}, \mathrm{d} A^{[1]}\rangle 
    +2i\langle \mathrm{d} A^{[1]},A^{[1]},A^{[1]}\rangle\nonumber\\&+2i\langle \mathrm{d} A^{[1]},A^{[3]},A^{[1]}\rangle+2i\langle \mathrm{d} A^{[1]}, A^{[1]},A^{[3]}\rangle+2i\langle \mathrm{d} A^{[1]},A^{[3]},A^{[3]}\rangle\,.\nonumber
\end{align}
After rescaling $A\to 2 A$ eqn. \eqref{eq:Sminusinfty} reproduces the familiar cubic Yang-Mills interaction plus non-local interactions originating from the last line in \eqref{eq:Sminusinfty}. \ivo{which has no projection on $\delta A^{[3]}$}
     
\end{comment}

\subsection{Associativity}\label{sec:ass}
Let us now study the properties of the product induced by the cubic vertex of the action. We begin with the $Q$-compatibility of the cubic vertex. For this we consider in \eqref{eq:S3mod} 
\begin{align}
     &\langle (Q \mathcal{V}_1({s_{-\infty}}))  e^{iF_{s_0}} \mathcal{V}_2(s_{0}) e^{-iF_{s_0}}  \mathcal{V}_3({s_{\infty}}) \rangle+\langle  \mathcal{V}_1({s_{-\infty}})  e^{iF_{s_0}} (Q \mathcal{V}_2(s_{0})) e^{-iF_{s_0}}  \mathcal{V}_3({s_{\infty}}) \rangle\rangle\\&+\langle  \mathcal{V}_1({s_{-\infty}})  e^{iF_{s_0}} \mathcal{V}_2(s_{0}) e^{-iF_{s_0}}  (Q \mathcal{V}_3({s_{\infty}})) \rangle=\mathrm{d}\eta\partial_\eta\;\langle  c(s_{-\infty}) \mathcal{V}_1({s_{-\infty}}) e^{iF_{s_0}} \mathcal{V}_2(s_{0}) e^{-iF_{s_0}}  \mathcal{V}_3({s_{\infty}}) \rangle\nonumber\,,
\end{align}
where the identities hold inside the integral. The adjoint action of $F_{s_0}$ on $Q$ produces the result by the chain map property. Now, since $\mathbb{R}^{(0|1)}$ has no boundary the $Q$-compatibility follows then from (the super version of) Stokes' theorem. This, in turn, confirms gauge invariance up to the second order. 

Next, we consider the associativity of the product induced by the cubic vertex. For this, we define the product of two vertices in picture $-1$ as
\begin{align}\label{m_2N=1}
    m_2(\mathcal{V}_1,\mathcal{V}_2)= c\,e^{iF} V_1\delta(\gamma) e^{-iF}\; V_2\delta(\gamma)\,.
\end{align}
Then we have 
\begin{align}
    m_2(\mathcal{V}_1 , m_2(\mathcal{V}_2,\mathcal{V}_3)) = c\,e^{iF_1} V_1\delta(\gamma) e^{-iF_1}c\,e^{iF_0} V_2\delta(\gamma) e^{-iF_0}\; V_3\delta(\gamma)\,,
\end{align}
which vanishes due to $c^2=0$, since \eqref{eq:fvdgf} contains no $b$-ghost. On the other hand, we have 
\begin{align}\label{eq:m2m2}
    m_2(m_2(\mathcal{V}_1,\mathcal{V}_2),\mathcal{V}_3))&= c\,e^{iF_1}c\,e^{iF_0}  V_1\delta(\gamma) e^{-iF_0}V_2 \delta(\gamma)e^{-iF_1} \; V_3\delta(\gamma)\nonumber\\ 
    &=  c\,2f_1(\frac{\mathfrak{f}_1}{2}-\gamma)e^{iF_0}  V_1\delta(\gamma) e^{-iF_0}V_2\delta(\gamma-\mathfrak{f}_1) \; V_3\delta(\gamma)  \,,
\end{align}
with $e^{iF_0}  V_1\delta(\gamma) e^{-iF_0}$ as in \eqref{eq:fvdgf}. After integration over\footnote{Recall that $\mathcal{V}_{-\infty}$ does not depend on $\gamma$.} $\gamma$
we end up with 
\begin{align}\label{eq:m2m2}
   \int \mathrm{d}\gamma\, m_2(m_2(\mathcal{V}_1,\mathcal{V}_2),\mathcal{V}_3)) & =- c\,f_1 f_0  
   (i(\mathrm{d} A_1+\delta A_1)\delta(-\mathfrak{f}_0)+B_1\delta(-\mathfrak{f}_0))\delta(-\mathfrak{f}_1)
   (C_2  A_3+ A_2 C_3)  \,,
\end{align}
which does not vanish after contraction with $V_{-\infty}$. We thus expect there to be a 3-product as well in order to restore gauge-invariance. 

\subsection{Geometric decomposition of $\mathcal{M}$}

We can try to give a geometric interpretation for the non-vanishing associator. Let us represent graphically the associator as

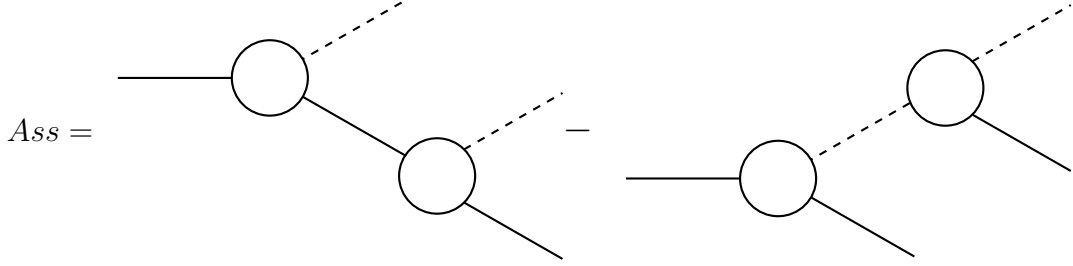
\begin{center}
\begin{minipage}{0.1\textwidth}
    \centering
    {\large \(Ass =\)} % Centered symbol at natural size
\end{minipage}
\begin{minipage}{0.45\textwidth} % Adjust width as needed
    \begin{tikzpicture}
        % Draw the first circle at the center with radius 0.5
        \draw[thick] (0,0) circle (0.5);

        % Line exiting to the left from the first circle
        \draw[thick] (-2,0) -- (-0.5,0);

        % Incoming line from the right (upward, at a 30° angle)
        \draw[thick,dashed] (30:2) -- (30:0.5); % Line inclined at 30° upwards
        
        % Incoming line from the right (downward, dashed and at a -30° angle)
        \draw[thick] (-30:2.07) -- (-30:0.5); % Line inclined at -30° downwards

        % Draw the second circle at the free end of the dashed line
        \draw[thick] (2.2,-1.3) circle (0.5); % Positioned at the free end of the dashed line

        % Lines exiting from the second circle to the right
        % Point on the edge of the second circle for the line inclined at +30°
        \draw[thick,dashed] (2.55,-0.95) -- ++(30:1.5); 
        % Point on the edge of the second circle for the line inclined at -30°
        \draw[thick] (2.55,-1.65) -- ++(-30:1.5); 
    \end{tikzpicture}
\end{minipage}
\hspace{-0.08\textwidth}
\begin{minipage}{0.03\textwidth}
    \centering
    {\Large \(-\)} % Centered symbol at natural size
\end{minipage}
\hspace{0.01\textwidth}
\begin{minipage}{0.45\textwidth} % Adjust width as needed
    \begin{tikzpicture}
        % Draw the first circle at the center with radius 0.5
        \draw[thick] (0,0) circle (0.5);

        % Line exiting to the left from the first circle
        \draw[thick] (-2,0) -- (-0.5,0);

        % Incoming line from the right (upward, at a 30° angle)
        \draw[thick,dashed] (30:2) -- (30:0.5); % Line inclined at 30° upwards
        
        % Incoming line from the right (downward, dashed and at a -30° angle)
        \draw[thick] (-30:2.07) -- (-30:0.5); % Line inclined at -30° downwards

        % Draw the second circle at the free end of the dashed line
        \draw[thick] (2.2,1.2) circle (0.5); % Positioned at the free end of the dashed line

        % Lines exiting from the second circle to the right
        % Point on the edge of the second circle for the line inclined at +30°
        \draw[thick,dashed] (2.55,1.55) -- ++(30:1.5); 
        % Point on the edge of the second circle for the line inclined at -30°
        \draw[thick] (2.55,0.85) -- ++(-30:1.5); 
    \end{tikzpicture}
\end{minipage}
\end{center}
\begin{comment}

\begin{center}
\begin{minipage}{0.1\textwidth}
    \centering
    {\large \(Ass =\)}
\end{minipage}
\begin{minipage}{0.45\textwidth}
\begin{tikzpicture}
        \draw[thick,dashed] (0,0) -- (2,-2);
        \draw[thick,dashed] (2,0) -- (3,-1);
        \draw[thick] (4,0) -- (2,-2);
        \draw[thick] (2,-2) -- (3,-3);
\end{tikzpicture}
\end{minipage}
\hspace{-0.19\textwidth}
\begin{minipage}{0.03\textwidth}
    \centering
    {\Large \(-\)}
\end{minipage}
\hspace{0.01\textwidth}
\begin{minipage}{0.45\textwidth}
    \begin{tikzpicture}
        \draw[thick,dashed] (0,0) -- (1,-1);
        \draw[thick] (2,0) -- (1,-1);
        \draw[thick,dashed] (1,-1) -- (2,-2);
        \draw[thick] (4,0) -- (2,-2);
        \draw[thick] (2,-2) -- (3,-3);
\end{tikzpicture}
\end{minipage}
\end{center}
\end{comment}
where the dashed lines represent picture-changed inputs and the solid lines represent untouched ones. The picture-changing operation is dictated by the odd vector field $v_f$, which translates the $(s,\theta)$ coordinates on the world-line as (recall Eq. \eqref{vector fields})
\begin{align}
    s & \mapsto s - \theta \eta \ , \nonumber\\
    \theta & \mapsto \theta + \eta \ ,
\end{align}
$\eta$ being the (odd) coordinate of the modulus. From the first diagram of the associator, we can read that the first and second inputs are shifted separately, thus leading to
\begin{align}
    s & \mapsto s - \theta \eta^1 - \theta \eta^2 \ ,\nonumber \\
    \theta & \mapsto \theta + \eta^1 + \eta^2 \ .
\end{align}
On the other hand, from the second diagram, we see that the first input is picture-changed twice, i.e.
\begin{align}
    s & \mapsto s - \theta \eta^1 \mapsto s - \theta \eta^2 - ( \theta + \eta^2 ) \eta^1 = s - \theta \eta^1 - \theta \eta^2 + \eta^1 \eta^2 \ ,\nonumber \\
    \theta & \mapsto \theta + \eta^1 \mapsto \theta + \eta^1 + \eta^2 \ .
\end{align}
The difference encoded in the extra term $\eta^1 \eta^2$ reflects the non-associativity of the operation involved which breaks gauge invariance, thus requiring the introduction of a higher term in the action. See also \cite{Ohmori:2017wtx} for a related discussion.

\subsection{Quartic term}\label{sec:quart}

Let us now include a quartic vertex of the form
\begin{align}
 C_4(\mathcal{V}_{-\infty},\mathcal{V}_{0},\mathcal{V}_{\epsilon},\mathcal{V}_{\infty})   &=\langle \mathcal{V}_{-\infty}(-\infty) \; e^{iF_{0}}  \mathcal{V}_0({0}) e^{-iF_{0}} \; e^{i(G_\tau+ F_{\tau})} \mathcal{V}_\epsilon(\epsilon) e^{-i(G_\tau+F_{\tau})} \;\mathcal{V}_\infty({{\infty}}) \rangle\,.
\end{align}
Here we relabeled $(\mathcal{V}_1 , \mathcal{V}_2 , \mathcal{V}_3)\to (\mathcal{V}_{-\infty} , \mathcal{V}_0 , \mathcal{V}_{\infty})$ for convenience. The integral over the zero-modes of $x,c$ and $\gamma$ is understood. With respect to the cubic term, we included the additional vertex operator 
\begin{align}
   \mathcal{V}_\epsilon(\epsilon) &= e^{i\epsilon H} 
 V(x,\psi,\beta ,c) \delta(\gamma ) e^{-i\epsilon H} \ ,
\end{align}
and 
\begin{align}
    G_\tau(\tau) = g_\tau(\tau)H+\mathfrak{g}_\tau(\tau)b \, , \qquad \mathfrak{g}_\tau(s) = \frac{\partial g(s)}{\partial \tau} \mathrm{d}\tau\, .
\end{align}
We choose the function $g_\tau=\tau \Phi(s)$, where $\Phi(s)$ is a differentiable function that takes the value 1 in the interval $[0,\epsilon]$. By acting with $Q$, we have
\begin{align}
%&Q\circ  C_4(\mathcal{V}_{-\infty},\mathcal{V}_{0},\mathcal{V}_{\tau},\mathcal{V}_{\infty})\equiv\nonumber\\
& C_4(Q\mathcal{V}_{-\infty},\mathcal{V}_{0},\mathcal{V}_{\epsilon},\mathcal{V}_{\infty}) + C_4(\mathcal{V}_{-\infty},Q\mathcal{V}_{0},\mathcal{V}_{\epsilon},\mathcal{V}_{\infty})+C_4(\mathcal{V}_{-\infty},\mathcal{V}_{0},Q\mathcal{V}_{\epsilon},\mathcal{V}_{\infty})+ 
C_4(\mathcal{V}_{-\infty},\mathcal{V}_{0},\mathcal{V}_{\epsilon},Q\mathcal{V}_{\infty})\nonumber\\
&\qquad\qquad\qquad\qquad =\left(\mathrm{d}\eta_\tau \partial_{\eta_\tau} +\mathrm{d}\eta_0 \partial_{\eta_0}+\mathrm{d}\tau\partial_\tau \right) C_4(\mathcal{V}_{-\infty},\mathcal{V}_{0},\mathcal{V}_{\epsilon},\mathcal{V}_{\infty}) 
 \end{align}
The $\eta$-derivatives give a vanishing contribution because of Stokes' theorem, as for $C_3$. The $\tau$-derivative pulls back to the boundaries at $\tau=0$ and $\tau=\epsilon$ instead. The former gives a vanishing contribution due to $\delta(\gamma)^2=0$. This leaves us with 
\begin{align}\label{eq:qm3}
 \langle \mathcal{V}_{-\infty}(-\infty)  \;e^{iF_{0}}  \mathcal{V}_0({0}) e^{-iF_{0}} \; e^{iF_{\epsilon}} \mathcal{V}_\epsilon(\epsilon) e^{-iF_{\epsilon}}\; \mathcal{V}_\infty({{\infty}}) \rangle\,.
\end{align}
Now, since $\mathcal{V}_\epsilon(\epsilon)$ is proportional to the $c$-ghost zero-mode, $F_\epsilon$, in addition to shifting the argument of $\delta(\gamma)$, it reduces to the supersymmetry transformation of this $c$-ghost. Therefore \eqref{eq:qm3} is identical to \eqref{eq:m2m2} thus establishing gauge in variance to third order. 

On the other hand, in the limit when\footnote{We consider this limit since it is sufficient for gauge invariance, otherwise this vertex would give a non-local contribution in the space-time action. } $\epsilon\to 0$, and modulo boundary terms in space-time, the non-vanishing contribution will come from\footnote{Double picture changing at $0$ or $1$ does not contribute since that results in $\delta(\gamma)^2$ from the other punctures.}
\begin{align}\label{eq:4newe}
 C^{(0,1)}_4 &=\#\int \mathrm{d}\eta_0 \mathrm{d}(\mathrm{d}\eta_0) \mathrm{d}\eta_1 \mathrm{d}(\mathrm{d}\eta_1)\eta_0\eta_1\langle V_{-\infty} c \;  V_0 \delta(\gamma-\mathfrak{f}_0) [q^2,\,\; e^{iG_1} V_\epsilon \delta(\gamma-\mathfrak{f}_1) e^{-iG_1}] \;V_\infty \delta(\gamma) \rangle\,,
\end{align}
where the coefficient $\#$ does depend on the parametrisations of $F_0$ and $F_\epsilon$ in terms of the two odd coordinates. We will simply fix the coefficient to be compatible with the gauge invariance of the action. 
%of the action. iplus terms that do not contribute a $\tau$-derivative. If add to this the correlator with picture changing at infinity\footnote{} we get a second contribution 
%\begin{align}\label{eq:4newe1i}
% C^{(1,\infty)}_4 &=\int \mathrm{d}\eta_1 \mathrm{d}(\mathrm{d}\eta_1) \mathrm{d}\eta_\infty \mathrm{d}(\mathrm{d}\eta_\infty)\langle \mathcal{V}_{-\infty} \;  \mathcal{V}_0 e^{iG_1} \mathcal{V}_1 e^{-iG_1} \;(1-(\eta_1 q)(\eta_\infty q))\mathcal{V}_\infty \rangle
%\end{align}
After integration over moduli space this results in a term proportional to %\footnote{There is a boundary at $\tau=\epsilon$ as well as at $\tau=0$. At $\tau=0$ we will have a $\delta(\gamma)^2$ or $\delta(\gamma)\delta(\gamma-\mathfrak{f}_0)$ depending on which operation we do first. We set this term to to zero.}
\begin{align}\label{eq:4ptsr}
\langle A,A,A,A\rangle=\frac{1}{2}\langle A,[A,A],A\rangle+ \langle A,A^2,A\rangle\,,
    %S^{(4)}&=\# \langle A, \; A,\; A,\; A \rangle\,,
\end{align}
which does not reproduce the expected quartic term for Yang-Mills theory. This discrepancy can, however, be resolved by recalling that in string theory the $t$ and $s$-channel contributions come from different regions in moduli space. The $s$-channel has the graphical representation 
\begin{equation}\label{fig:wl3st}
\begin{gathered}\begin{tikzpicture}
\draw[very thick] (-3, 0) -- (-1, 0);
\draw[dashed] (-1, 0) -- (0, 0);
\draw[very thick] (0, 0) -- (3, 0);
\node at (-3, 0) {\textbullet};
\node at (0, 0) {\textbullet};
\node at (3, 0) {\textbullet};
\node at (-3, -0.4) {$\mathcal{V}_{-\infty}(-\infty)$ };
\node at (3, -0.4) {$\mathcal{V}_\infty(\infty) $ };
\node at (1, 1.6) {$\mathcal{V}_\epsilon(\epsilon)$ };
\node at (0, 1.6) {$\mathcal{V}_0(0)$ };
\draw[very thick] (0, 0) -- (1, 1);
\node at (1, 1) {\textbullet};
\draw[very thick] (-1, 0) -- (0, 1);
\node at (0, 1) {\textbullet};
\node at (-1, 0) {\textbullet};
\end{tikzpicture}\end{gathered} \ ,
\end{equation}
where the dashed line indicates the integral over the even modulus $\tau$. The $t$-channel, instead has the representation 
\begin{equation}\label{fig:wl3stt}
\begin{gathered}\begin{tikzpicture}
\draw[very thick] (-3, 0) -- (0, 0);
\draw[dashed] (0, 0) -- (0, 1);
\draw[very thick] (0, 0) -- (3, 0);
\node at (-3, 0) {\textbullet};
\node at (0, 0) {\textbullet};
\node at (3, 0) {\textbullet};
\node at (-3, -0.4) {$\mathcal{V}_{-\infty}(-\infty)$ };
\node at (3, -0.4) {$\mathcal{V}_\infty(\infty) $ };
\node at (1, 2.6) {$\mathcal{V}_\epsilon(\epsilon)$ };
\node at (0, 2.6) {$\mathcal{V}_0(0)$ };
\draw[very thick] (0, 1) -- (1, 2);
\node at (1, 2) {\textbullet};
\node at (0, 2) {\textbullet};
\draw[very thick] (0, 1) -- (0, 2);
\node at (0, 1) {\textbullet};
\end{tikzpicture}\end{gathered} \ .
\end{equation}
The $t$-channel thus corresponds to a cyclic permutation of the $s$-channel followed by a permutation of $\mathcal{V}_1$ and $\mathcal{V}_3$. On the time-ordered world-line, the latter operation amounts to moving $s_\infty$ to the left of $s_{-\infty}$ which implies a relative sign in the fermion propagator. Thus, summing over the $s$ and $t$ channels amounts to antisymmetrizing in the $in$ and $out$ states. Then the r.h.s. of \eqref{eq:4ptsr} is replaced by
\begin{align}\label{eq:4ptsra}
\langle A,A,A,A\rangle=\langle A,[A,A],A\rangle=-\frac{1}{2}\langle [A,A], [A,A]\rangle \,.
\end{align}
Combining this with $S^{(2)}+S^{(3)}$ we recover the vertices of the perturbative Yang-Mills action for $A^{[1]}$, plus non-local modifications induced by $A^{[3]}$ which is in turn sourced by $A^{[1]}$ through \eqref{eq:19-12-2}. 

We want to remark that there are no local vertices of higher order since each additional puncture will come with one extra bosonic modulus and one extra fermionic modulus. However, to produce a total derivative in $\tau$, two fermionic moduli are needed as shown in Eq. \eqref{eq:4newe}. The absence of higher-order vertices then follows from this simple counting argument.

We note in closing that, while gauge-invariance of the action requires a quartic (contact) term, the full Yang-Mills equations of motions are already implied at cubic order upon elimination of the Lagrange multiplier $B^{[\bullet]}$. This is consistent as the quartic vertex does not depend on $B^{[\bullet]}$.

\subsection{Non-linear gauge transformation}

In this section, we consider the symmetrisation of the definition of the cubic term; this leads to the correct expression of the non-linear gauge transformation of the gauge field $A^{[1]}$.

Let us consider in the path integral the case where the ghost $c$ and $\delta ( \gamma )$ are inserted with a different prescription. In particular, let us consider
\begin{align}\label{eq:V123otherprescription} 
 \mathcal{V}_{-\infty}(s_{-\infty})&=e^{i(s_{-\infty})H} V_{-\infty}(x(s_{-\infty}),\psi(s_{-\infty}),\beta(s_{-\infty}),c(s_{-\infty}))  
\delta ( \gamma ( s_{-\infty} ) ) \;e^{-i(s_{-\infty})H}\nonumber\\
       \mathcal{V}_0(s_0)&=e^{i(s_0)H} V_0(x(s_0),\psi(s_0),\beta(s_0),c(s_0))\delta(\gamma(s_0))e^{-i(s_{0})H}\nonumber\\
        \mathcal{V}_{\infty}(s_{\infty})&=e^{i(s_{\infty})H} c ( s_\infty )  V_{\infty}(x(s_{\infty}),\psi(s_{\infty}),\beta(s_{\infty}),c(s_\infty)) e^{-i(s_{\infty})H}\,,
\end{align}
where the vertex in $s_0$ is picture-changed. The cubic term thus reads
\begin{align}\label{eq:S3notherprescription}
    S^{(3)}_2 & = \int d^4 x \left( 
    - i \left \langle A , d A + \delta A - i B, A \right\rangle - \left\langle C , A^* , A \right\rangle \right) \ .
\end{align}
If we take the difference between \eqref{eq:S3n} and \eqref{eq:S3notherprescription} we obtain
\begin{align}
    S^{(3)} - S^{(3)}_2 & = \int d^4 x \left( 
    2 i \left \langle A , d A + \delta A - i B, A \right\rangle + \left\langle A , A^* , C \right\rangle + \left\langle C , A^* , A \right\rangle \right) \ .
\end{align}
The second and third terms sum to the usual non-linear term of the gauge transformation for $A$; in particular, if we consider the term with $A^{(1)}, A^{*(1)}$ and $C^{(0)}$, we can contract the $\psi$ of the two one-forms thus getting a commutator
\begin{align}
    \left\langle A_{\mu}^a T^a \psi^\mu (-\infty) A^{*b}_\nu T^b \psi^\nu (0) C^c T^c + C^a T^a A^{*b}_\nu T^b \psi^\nu ( 0 ) A_\mu^c T^c \psi^\mu ( \infty ) \right\rangle = \\
    = \langle \frac{i}{2} A^{*\mu b} \left( A_\mu^a C^c - C^c A_\mu^a \right) T^a T^b T^c \rangle = \frac{i}{2} \left\langle A^*_\mu , \left[ A^\mu , C \right] \right\rangle \ .
\end{align}

\section{Background field description}\label{sec:background fields}
In this section, we will consider a different approach by treating the insertion at $s=0$ and $\epsilon$ not as states but rather as background fields in which the world-line propagates. In particular, they are not the result of the mapping of a state to these punctures but rather external fields in which the \textit{in} state (inserted at $\infty$) propagates. This, however, raises several questions. The first concerns the definition of an \textit{in} state in an arbitrary background field. A meaningful definition is given by the cohomology of a BRST-differential, which requires $Q$ to be nilpotent. We will see that this is the case when the equations of motion are satisfied. Furthermore, there is no notion of a decomposition of moduli space for background fields. The notion of moduli space is still well defined: When expanding the world-line path integral in powers of the background fields this naturally amounts to integrating over punctures modulo gauge redundancies and automorphisms, but there is no notion of vertices (maps between states) and propagators. As a result, the space-time functional of the background fields, obtained by simply integrating over an arbitrary number of bosonic moduli $\tau_i$, has no a priori reason to be a local functional. In string theory, integration over bosonic moduli comes with inverse powers of $\alpha'$ so that in the infinite tension limit this functional should be local. On the world-line, there is no notion of large tensions but we can still make the local approximation. At first order in the background field, the result is local as the 3-punctured line has no bosonic moduli. As we saw, gauge-invariance requires a quartic term in addition, which is still local in the bosonic direction. We will show in this section that the resulting space-time functional has just the right properties to represent a nilpotent BRST operator in a background field. 
%%%%%%%%

In the quadratic approximation, there is no difference to section \ref{sec:perturbative_action} since there are neither even nor odd moduli to be integrated over. Note, however, that if the insertions at $\pm\infty$ are treated as background fields there will be extra terms in the form of total derivatives since $Q$ will act via the adjoint action rather than the left action (as on states). As a result, there are extra terms containing $p$ that act as total derivatives. 

The cubic contribution is again given by \eqref{eq:S3n}, up to total derivatives, except that now we can not trade the anticommutator $\{p,\cdot A\}$ in favour of $\delta A$ as we did in \eqref{eq:S3ns}. Indeed, from \eqref{eq:two-points} it follows that contraction with $\{p,\cdot A\}$ results in $-i\partial_x$ on the functions on the left of $p$ and in $i\partial_x$ on the functions on the right (the change in sign is due to the sign function resulting from the $p,x$ contraction). This is precisely the opposite of a total derivative. Thus, up to cubic order the space-time functional becomes
\begin{align}\label{eq:19-12-1-2}
  S^{(2)}+S^{(3)}=&  \langle A,-\Box A+ i\mathrm{d} B +i\delta B\rangle+i\langle A, \mathrm{d} A +i\{p,\cdot A\}-i B,A\rangle\\
    =& -\langle \mathrm{d} A, \mathrm{d} A \rangle -\langle \delta A, \delta A\rangle +i\langle \mathrm{d} A, B\rangle +i \langle \delta A , B\rangle+i\langle A, \mathrm{d} A +i\{p,\cdot A\}-i B,A\rangle\nonumber
\end{align}
where again, we focused fields of ghost number $0$ only. 

Alternatively, if we write \eqref{eq:19-12-1-2} as
\begin{align}\label{eq:19-12-3}
    S^{(2)} + S^{(3)} = & \langle A , p^2 - \lbrace p , \cdot A \rbrace + i \mathrm{d} A , A \rangle + \langle i \mathrm{d} A + i \delta A , B \rangle + \langle A , B , A \rangle \ .
\end{align}

where\footnote{We have $(\mathrm{d} A^{[1]})_{\mu\nu}=\frac{1}{2}\left(\partial_\mu A_\nu- \partial_\nu A_\mu\right)$} 
\begin{align}
    H(A)=(p^2-\{p,\cdot A\}
&+i\mathrm{d}A)%+\gamma(p\cdot\psi+A)+b\gamma^2
\,,\qquad \mathrm{d}A=(\mathrm{d}A)_{\mu\nu}\psi^\mu\psi^\nu\,,
\end{align}
reproduces the expected BRST Hamiltonian in a background field, up to cubic order. Note that 
\begin{align}
    Q(A)= -c(p^2-\{p,\cdot A\} +i\mathrm{d}A)+\gamma(p\cdot\psi-A)+b\gamma^2
\end{align}
is nilpotent without assuming equations of motion for $A$, but the latter are still enforced by variation w.r.t. $B^{[\bullet]}$, which implies 
\begin{align}\label{eq:19-12-2-2}
i\delta A^{[1]}+\langle A, A\rangle&=0\nonumber\\
\langle\cdot, i \mathrm{d} A^{[3]}\rangle + \langle\cdot,A^{[3]}, A^{[3]}\rangle&=0\nonumber\\
\langle\cdot,i \mathrm{d} A^{[1]}\rangle +\langle\cdot,i\delta A^{[3]}\rangle +\langle\cdot,A,A\rangle&=0\,.
\end{align}
The last term on the third line is of the form 
\begin{align}
  \langle\cdot,A,A\rangle= \langle\cdot, \frac{1}{2}[A^{[1]}_\mu,A^{[1]}_\nu]\psi^\mu\psi^\nu \rangle +\cdots \ ,
\end{align}
where $\cdots$ stands for terms that involve $A^{[3]}$. As in section \ref{sec:perturbative_action}, acting with $\delta$ on the third equation then implies the Yang-Mills equation up to non-local modifications.

The discussion of $Q$-compatibility and associativity in section \ref{sec:ass} carries over to the present case if we replace the left action of $Q$ by the adjoint action and accordingly replace $\delta A$ by $\{p,\cdot A\}$ in eqn. \eqref{eq:m2m2}. Similarly, the construction of the quartic vertex carries over from section \ref{sec:quart} except that now the $t$-channel contribution is absent. We represent the background fields by just inserting a fat point, $\bullet$ on the world-line as in \eqref{fig:wl3sonly}, rather than an extra line corresponding to a state: 
\begin{equation}\label{fig:wl3sonly}
\begin{gathered}\begin{tikzpicture}
\draw[very thick] (-3, 0) -- (-1, 0);
\draw[dashed] (-1, 0) -- (0, 0);
\draw[very thick] (0, 0) -- (3, 0);
\node at (-3, 0) {\textbullet};
\node at (0, 0) {\textbullet};
\node at (3, 0) {\textbullet};
\node at (-3, -0.4) {$\mathcal{V}_{-\infty}(-\infty)$ };
\node at (3, -0.4) {$\mathcal{V}_\infty(\infty) $ };
\node at (0, -0.4) {$\mathcal{V}_\epsilon(\epsilon)$ };
\node at (-1, -0.4) {$\mathcal{V}_0(0)$ };
\node at (-1, 0) {\textbullet};
\end{tikzpicture}\end{gathered} \ .
\end{equation}
Accordingly, the quartic contribution is now given by just \eqref{eq:4ptsr}, that is 
\begin{align}\label{eq:4ptn0sr}
\langle A,A,A,A\rangle=\frac{1}{2}\langle A,[A,A],A\rangle+ \langle A,A^2,A\rangle\,,
    %S^{(4)}&=\# \langle A, \; A,\; A,\; A \rangle\,.
\end{align}
Adding this to \eqref{eq:19-12-3}, $Q(A)$ is modified to 
\begin{align}
    Q(A)= -c\left[(p-\hat A)^2+\frac{i}{2} F(A)\right]+\gamma\psi\cdot (p-\hat A)+b\gamma^2\,,\qquad F(A)=2\mathrm{d}A-i[A,A]\,.
\end{align}
and $A=\hat A_\mu\psi^\mu$. This is now the complete local correction to the BRST operator in a background field. What we have thus found is that the integration over the odd directions of moduli space are encoded in a specific deformation of the Hamilton $H(A)$ by background fields.  

One may wonder what is the role of the supercharge
\begin{align}
    q(A)= \psi\cdot (p-\hat A) \ ,
\end{align}
which, in some sense is a more fundamental object since $H=\frac{1}{2}\{q,q\}$ is derived from it. One way to interpret $q(A)$ is that it provides the relation between background fields and states through 
\begin{align}
  \mathcal{V}_\infty=  -Q(A)\delta'(\gamma(\infty))
\end{align}
so that an equivalent formula for $S=S^{(2)}+S^{(3)}+S^{(4)}$ is given by 
\begin{align}
    -\frac12\int \mathrm{d} c\, \mathrm{d}\gamma \;\langle A, Q(A), Q(A) \delta'(\gamma)\rangle 
\end{align}
It would be nice to repeat this with the puncture at $-\infty$. However, since that puncture is in the polynomial representation at picture zero, this would necessitate a $\gamma^{-1}$ instead of  $\delta'(\gamma)$. This is, however not possible in the small Hilbert space \cite{Belopolsky:1997jz, Witten:2012bg}. In string theory, this is different since both punctures are in picture $-1$. For the point particle, this can be achieved by embedding the present model in a world-line with $\mathcal{N}=2$ world-line supersymmetry. We will return to this in a future publication.  

\section{Conclusions}

In this paper, we have derived the space-time action for the $\mathcal{N}=1$ supersymmetric world-line following the procedure of string field theory. Concretely, we use the BRST quantization of the Hamiltonian world-line action, extended to supermoduli space, to construct a functional of the space-time BV multiplet as an integral form on moduli space. This results in a Yang-Mills theory, modified by non-local interactions, arising from the elimination of massless auxiliary $3$-form potentials that would be massive in string theory. Space-time SUSY does not feature in this model. 

We found that, if the world-line punctures are identified with states, the space-time functional we construct has the interpretation as a perturbative expansion of the action for which we give the corresponding geometric decomposition of super moduli space $\mathcal{M}$. If the world-line punctures are identified with background fields (as in background-independent string field theory), the space-time functional is naturally related to the space-time BV differential in a background field. For the latter, there is no decomposition of $\mathcal{M}$ but in either case, a simple counting of dimensions on $\mathcal{M}$ shows that there are no local vertices of degree order higher than four. The latter approach also provides a justification of the approach followed in \cite{Dai:2008bh,Bonezzi:2018box,Bonezzi:2020jjq} where one parametrizes the BRST-differential in terms of local fields and derives equations of motions for the latter from nilpotency (see \cite{Grigoriev:2021bes} for a general description). We aim to give more details on this connection in forthcoming work. It would be interesting to have related results, both concerning the higher order vertices in open superstring field theory as well as regarding the construction of the BRST-differential in a background field for string theory. In \cite{Dai:2008bh,Bonezzi:2018box,Bonezzi:2020jjq}, the supercharge and the Hamiltonian of the BRST-differential play separate roles, as operator-state maps and as a projection on the bosonic moduli part of $\mathcal{M}$ which is now manifest. 
 
We have not explored the 2-form theory arising by considering a BV-multiplet of even parity instead. This should be a straightforward but interesting extension.

\section*{Acknowledgments}
\label{sec:Aknowledgements}
%%%%%%%%%%%%%%%%%%%%%%%%%%%%%%%%%%%%%%%%%%%%%%%%%%%%%%%%%%%%%

 This work was funded by the Excellence Cluster Origins of the DFG under Germany’s Excellence Strategy EXC-2094 390783311.

\appendix

\section{More details on picture changing}
In this appendix, we give more details on the computation leading up to Eq. \eqref{eq:fvdgf}. Let us start from the expression of the PCO, given by the adjoint action of $e^{iF}$. By using the Baker-Campbell-Hausdorff formula, we can decompose it as
\begin{align}
    e^{iF}=e^{-fp\cdot \psi}e^{- f 2 b\gamma}  e^{- \mathfrak{f} \beta} e^{f \mathfrak{f}b}  \,,
 \end{align} 
where $\mathfrak{f}= f'\mathrm{d}\eta$. We then find 
%My computation is:
\begin{align}
    e^{iF} V \delta ( \gamma ) e^{-iF} & = \left[ 1 - f ( p \psi + 2 b \gamma - \mathfrak{f} b ) \right] V \delta ( \gamma - \mathfrak{f} ) \left[ 1 + f ( p \psi + 2 b \gamma - \mathfrak{f} b ) \right] \ ,
\end{align}
where $f,\mathfrak{f}$ and $\gamma$ act multiplicatively, while $b$ and $\beta$ act as derivations (thus the shift in the argument of $\delta ( \gamma )$). Let us focus on the $p \psi$ term. Suppose you have
\begin{align}
    p_\mu \psi^\mu A_\nu \psi^\nu = \psi^\mu \left[ \left( p_\mu A_\nu - A_\nu p_\mu \right) + A_\nu p_\mu \right] \psi^\nu \ .
\end{align}
The first term in round brackets reads
\begin{align}
    \psi^\mu \left[ p_\mu , A_\nu \right] \psi^\nu = \psi^\mu \left[ p_\mu , x_\rho \right] \partial^\rho A_\nu \psi^\nu = - i \partial_\mu A_\nu \psi^\mu \psi^\nu = -i ( \mathrm{d} + \delta ) A^{[1]} \ .
\end{align}
The other term goes together with the other piece coming from the adjoint action of $F$. In particular, we can write
\begin{align}
    & - f p_\mu \psi^\mu A_\nu \psi^\nu \delta ( \gamma - \mathfrak{f} ) + A_\nu \psi^\nu \delta ( \gamma - \mathfrak{f} ) f p_\mu \psi^\mu = \nonumber\\
    & = f \left[ i ( \mathrm{d} + \delta ) A \delta ( \gamma - \mathfrak{f} ) - A_\nu p_\mu \psi^\mu \psi^\nu \delta ( \gamma - \mathfrak{f} ) - A_\nu p_\mu \psi^\nu \psi^\mu \delta ( \gamma - \mathfrak{f} ) \right] = \nonumber\\
    & = f \left[ i ( \mathrm{d} + \delta ) A \delta ( \gamma - \mathfrak{f} ) - A_\nu p_\mu \left( \psi^\mu \psi^\nu + \psi^\nu \psi^\mu \right) \delta ( \gamma - \mathfrak{f} ) \right] = \nonumber \\
    &= f \left[ i ( \mathrm{d} + \delta ) A \delta ( \gamma - \mathfrak{f} ) - 2 A\cdot p \delta ( \gamma - \mathfrak{f} ) \right] 
\end{align}
The adjoint action of $e^{-f p \psi}$ can be schematically described as follows:
\begin{align}
    -f p_\mu \psi^\mu V \delta (\gamma) + V \delta (\gamma) p_\mu \psi^\mu = - f \left[ p_\mu \psi^\mu V + V p_\mu \psi^\mu \right] \delta (\gamma) = - f \left[ \psi^\mu \left( p_\mu V \right) + \left\lbrace V, \psi^\mu \right\rbrace p_\mu \right] \delta (\gamma)\nonumber\,,
\end{align}
where the first term leads to $\mathrm{d} + \delta$ and the second term leads to $2 V \cdot p$. We can thus rewrite
\begin{align}
    e^{iF} V (c, \psi) \delta(\gamma) e^{-iF} & = \left[ 1 - f p \psi \right] V ( c - 2 f \gamma + f \mathfrak{f} , \psi ) \delta ( \gamma - \mathfrak{f} ) \left[ 1 + f p \psi \right] = \nonumber\\ 
    & =  V \delta(\gamma-\mathfrak{f})+if( \mathrm{d A} + \delta A + 2 i A\cdot p)\delta(\gamma-\mathfrak{f})-2f(-\frac{\mathfrak{f}}{2}+\gamma)A^* \delta(\gamma-\mathfrak{f})\nonumber \\
    & -i f c (\mathrm{d} A^* + \delta A^* + 2 i A^* \cdot p) \delta (\gamma - \mathfrak{f}) -2f( -\frac{\mathfrak{f}}{2}+\gamma)B\delta'(\gamma-\mathfrak{f})\nonumber \\
    & -ifc(\mathrm{d B}+\delta B + 2 iB\cdot p)\delta'(\gamma-\mathfrak{f}) + if (\mathrm{d} C + \delta C + 2 i C \cdot p) \delta' (\gamma - \mathfrak{f} ) \nonumber\\
    & -2 f (- \frac{\mathfrak{f}}{2} + \gamma ) E\delta^{(2)}(\gamma - \mathfrak{f} ) -i f c (\mathrm{d} E + \delta E + 2 i E \cdot p) \delta^{(2)} ( \gamma - \mathfrak{f} )\nonumber\nonumber \\
    & +i f (\mathrm{d} D + \delta D + 2 i D \cdot p ) \delta^{(2)}(\gamma-\mathfrak{f}) +\cdots \ .
\end{align}

\section{Symplectic detour}\label{sec:Symplectic detour}

In the discussion of the linearised equations of motion in Section \ref{sec:Path-integral quantization}, we have chosen a specific form of the vertex (with picture number 0, i.e., in the $\mathbf{1}_\gamma$ representation of the $\beta,\gamma$ system) in $-\infty$, that is $\mathcal{V} = \left( A + c A^* \right) \mathbf{1}_x \cdot \mathbf{1}_\gamma$. This choice reflects the fact that out of the fields contained in a picture $-1$ vertex (see \eqref{picture-1vertex}) we allow only for fields which can be obtained via picture changing. This choice manifestly breaks the non-degeneracy defined by the path integration leading to a \emph{pre-symplectic} form rather than a symplectic one: basically, any term proportional to $\beta$ has no partner to couple with. This fact simply reflects the choice of not introducing antifields (except for $A^*$) in our description, as we included only fields with non-negative ghost degrees (again, except for $A^*$). In the following, we briefly comment on the introduction of antifields in the system and the resulting symplectic extension (reflecting the full BV theory) of our model.

As we suggested above, let us consider the introduction of antifields with negative ghost degree; in particular, we modify the definition of the picture-zero vertex by also allowing for polynomials in $\gamma$ as
\begin{align}
    \mathcal{V} = \left( A + c A^* + c \gamma C^* + \gamma B^* + c \gamma^2 D^* + \gamma^2 E^* + \ldots \right) \mathbf{1}_x \cdot \mathbf{1}_\gamma \ ,
\end{align}
where the notation $f^*$ for any field $f$ is used on purpose to identify field-antifield pairs. This completion lifts the pre-symplectic form to a symplectic one.\footnote{Note that we have implicitly imposed a constraint here: a priori one could have that all the fields and ghosts are included in the picture -1 vertex and the antifields and antighosts are in the picture 0 vertex. In particular, the field in $c A^* \delta ( \gamma )$ could be unrelated to the antifield of the gauge field $A$. This would lead to a BF-type model, as a consequence of the fact that fields and antifields live in different complexes (i.e., representations of the $\beta$ - $\gamma$ system).}
%\begin{align}
%    \omega &= \left\langle \delta (A + c A^* + c \gamma C^* + \ldots) , \delta \left( A + c A^* + \beta C + \ldots \right) \delta ( \gamma )\right\rangle = \nonumber \\
%    & = \int d^4x \left[ \langle \delta A , \delta A^* \rangle + \langle \delta C , \delta C^* \rangle + \langle \delta B , \delta B^* \rangle + \ldots \right] \ .
%\end{align}
The kinetic term would then read
\begin{align}
    S^{(2)} & = \langle \mathcal{V}_{-\infty}, Q \mathcal{V}_{\infty} \delta ( \gamma ) \rangle = \\
    & = \int d^4x \left[ \left\langle A , - \Box A + i (d + \delta ) B \right\rangle + \left\langle A^* , i ( d + \delta ) C + 2 E \right\rangle + \left\langle B^* , \Box C - 2 i ( d + \delta ) E \right\rangle + \ldots \right] \ , \nonumber
\end{align}
where the variation w.r.t. the antifields encodes the gauge transformations described in \eqref{gauge N=1}. The cubic interaction term is constructed by following the same prescription as in the previous sections (here, we assume to insert states at the punctures as in Sec. \ref{sec:perturbative_action}): we insert a $c$  in the vertex at $s = -\infty$, a $\delta ( \gamma )$ in the vertices at $s=0, + \infty$ and we picture-change the vertex at $s=0$. We have
\begin{align}
    S^{(3)} = \int d^4 x \Big[ & i \left\langle A , dA + \delta A - i B , A \right\rangle + \left\langle A , A^* , C \right\rangle \nonumber \\
    & - i \left\langle B^* , ( d A + \delta A ) - i B , C \right\rangle - 2 \left\langle B^* , A^* , D \right\rangle + \ldots \Big] \ ,
\end{align}
and analogously for the higher vertices. The reduction to the description given in the previous sections is done by setting the antifields (except for $A^*$) to zero or, analogously, by truncating to $\gamma = 0$ the polynomial expansion of the vertex in $-\infty$. This reduction is coisotropic, thus leading to a degeneration of the symplectic form when restricted to this subspace.

\bibliography{HSmaster}

\end{document}